\documentclass{iopart}
\usepackage{graphicx}

\begin{document}

\newcommand{\be}{\begin{equation}}
\newcommand{\ee}{\end{equation}}

\newcommand{\bea}{\begin{eqnarray}}
\newcommand{\eea}{\end{eqnarray}}

\newcommand{\azAngle}{\varphi}
\newcommand{\EctSub}{{\rm CT}}

\newcommand{\height}{6.75cm}
\newcommand{\width}{6cm}
\newcommand{\picangle}{270}
\newcommand{\capwidth}{15.5cm}


\article{}{Quantum stabilization of Z-strings, a status 
  report on D=3+1 dimensions}  
\author{O.~Schr\"oder$^1$\footnote{talk presented at QFEXT'07 by
    O.S.}, N.~Graham$^2$, M.~Quandt$^3$, H.~Weigel$^4$}  
\address{$^1$ science+computing ag, Hagellocher Weg 73, 72070
  T\"ubingen, FRG\\ 
  $^2$ Department of Physics, Middlebury College, Middlebury, VT
  05753, USA\\
  $^3$ Institute for Theoretical Physics, T\"ubingen University, 72076
  T\"ubingen, FRG\\ 
  $^4$ Fachbereich Physik, Siegen University, 57068 Siegen, FRG} 
            \ead{: $^1$oliver.schroeder@science-computing.de\\ 
  \hspace*{4.75em} $^2$ngraham@middlebury.edu\\ 
  \hspace*{4.75em} $^3$quandt@tphys.physik.uni-tuebingen.de\\ 
  \hspace*{4.75em} $^4$weigel@physik.uni-siegen.de} 
\date{\today}
\begin{abstract}
\noindent We investigate an extension to the phase shift formalism for
calculating one-loop determinants. This extension is motivated by
requirements of the computation of Z-string quantum energies in D=3+1
dimensions. A subtlety that seems to imply that the vacuum
polarization diagram in this formalism is (erroneously) finite is
thoroughly investigated.  
\end{abstract}

\section{Introduction and motivation\label{sec_intro}}
Z-strings were first discovered as solutions to the
classical field equations of the electroweak Standard Model
by Nambu \cite{Nambu} in the context
of bound pairs of magnetic monopoles. Later on they were rediscovered
--- as independent objects in their own right --- by Vachaspati
\cite{Vachaspati:1992fi}.

The main point of interest in our study of Z-strings is their
stability. If they are stable, they might be relevant for a variety of
reasons:
They would be the only static solitons in the Standard Model,
networks of Z-strings might play an important role in baryogenesis
\cite{Brandenberger}, and they would be a possible source of the
primordial magnetic field.   
For a general overview of applications and properties of Z-strings
along with a large collection of references cf. \cite{VachaspatiReview}. 

Because there are no configuration with nontrivial topology in
the electroweak model, stability of Z-strings is only possible on energetic
grounds. Classically, Z-strings are unstable for the physical values
of the parameters of the electroweak model. 
Since fermions bind strongly to the core of the Z-string, it may be possible
to achieve stability by occupying $N$ fermionic bound states that are
generated in the background of the Z-string so
that the the resulting total energy of the system is less
than the energy of $N$ free fermions. A consistent $\hbar$ expansion
then requires to include the contribution of the fermionic
determinant to the energy. 

In D=3+1 dimensions, renormalization issues make investigation of the
Z-string difficult since it is non-perturbative. Investigations
have either failed to draw convincing conclusions \cite{Groves:1999ks}
or focused on  D=2+1 dimensions \cite{ Schroeder:2006, Graham:2006}.
Here we will discuss attempts to solve the problems posed
by these earlier investigations. 

The structure of this paper is as follows: In section \ref{sec_tp},
we discuss our method for computing the
fermion determinant and the problems posed by the Z-string for our approach.
We discuss the possible solutions in D=2+1 dimensions and the
extension necessary for D=3+1 dimensions. In section
\ref{sec_fake} we investigate whether the extension proposed for the
Z-string works in the case of a simple boson model. Problems
involving
the second order Born approximations are discussed and resolved by
comparing different formulations of our approach.
In section \ref{sec_conclusions} we present some
conclusions and an outlook.

\section{Technical prerequisites\label{sec_tp}}
The Z-string has the following structure:
\bea 
\phi & = \biggl( \begin{array}{c} \phi_+\\ \phi_0   
                  \end{array} \biggr) 
       = v \biggl( \begin{array}{c} 0 \\   
                    f_H(\rho) e^{i \azAngle} \end{array} \biggr), \,
g \vec{Z}  =   \frac{\hat{\azAngle}}{\rho} 2
f_G(\rho), \label{eqZstring}
\eea
with $\phi$ the iso-spinor Higgs field and $\vec{Z}$ the only
non-vanishing component of the $SU(2) \times U(1)$ gauge
fields. As usual, $\rho$ and $\azAngle$ denote the radial
and angular coordinate in the plane perpendicular to the Z-string and
$v$ is the vacuum expectation value of the Higgs field.
The string configuration is defined in two spatial dimensions.
It is translationally invariant in any additional space coordinate,
which we call \emph{flat} dimension(s). Finite
classical energy (for D=2+1, or classical energy per unit length in
D=3+1) requires both $f_H \to 0$ and $f_G \to 0$ for $\rho \to 0 $.
Furthermore, the kinetic energy of the Higgs must fall off sufficiently fast
at large distances,
\be
| (\partial_{\mu} - i g Z_{\mu}) \phi |^2 =
\mathcal{O}(1/\rho^{2+\epsilon}), \epsilon > 0 \textrm{ for } \rho \to \infty.
\label{eq_fastfalloff}
\ee
In particular, this implies that $\phi^\dagger\phi = |f_H|^2 \to 1$
and $f_G \to 1$ at $\rho \to \infty$. The condition
(\ref{eq_fastfalloff}) clearly mixes different orders 
in perturbation theory (PT)
and already indicates that if one sticks to a \emph{fixed} order in
PT one will end up with IR divergent quantities.
In the systematic expansion of arbitrarily many fermion species and at
next to leading order in $\hbar$, the quantum correction to the energy,
i.e. the vacuum polarization energy, may be computed from the fermion
determinant in the background of the potential $V(\vec{x})$ that is
generated by the string. In turn, this determinant is given in terms
of scattering data from this potential \cite{Leipzig}. This
formulation is effective for renormalization because the n$^{th}$
order contribution in the Born expansion equals the Feynman diagram
for the fermion loop with n insertions of V. To render the integral
over scattering data finite, we subtract enough Born
terms and add them back in as Feynman diagrams. The renormalization of
the latter is adopted from perturbation theory and is standard. Hence
the vacuum polarization energy is given by
\bea 
\fl  E_{\rm vac} = \frac{1}{2} \sum_{b.s.} f_1(\omega^{b.s.}_j)
- \frac{1}{2 \pi}\int \rmd k\, f_2(k) \sum_{M} 
  \Big[\delta_{M}(k) - \sum_{n=1}^N \delta_{M}^{(n)}(k)
  \Big] \nonumber \\  +  \sum_{n=1}^N E_{FD}^{(n)} +
  E_\EctSub, \label{eqVacPolEn}
\eea
where $\omega_j^{b.s.}$ denotes the fermion bound state energies, $\delta_M$
the (full) phase shift in angular momentum channel $M$,
$\delta_{M}^{(n)}$ its n$^{th}$ Born approximation, $N$ denotes the
number of Born approximations necessary to render the momentum
integral finite, $E_{FD}^{(n)}$ is the
energy contribution computed from Feynman diagrams with n external
legs and $ E_\EctSub$ is the energy resulting from the counter terms.
The functions $f_{1,2}$ vary with the number of flat dimensions
\cite{GrahamInterface}. For the Z-string in D=2+1, the number of flat
dimensions is 0 and we 
obtain $f_1(\omega) = m - |\omega|, f_2(k) = -k/\sqrt{k^2+m^2}$. For D=3+1, the
number of flat dimensions is 1 and we obtain $f_1(\omega) =
(\omega^2(1+\ln{(\omega^2/m^2)})-m^2)/(4 \pi), f_2(k) = (k/2 \pi)
\ln{(1+k^2/m^2)}$. Of course, the Feynman diagram and counter term
contribution also differ for D=2+1 and D=3+1, but these expressions
are well-known and will not be repeated here. 
When computing the Born approximation for scattering in the
Z-string background, one encounters two problems:
The full scattering problem is decomposed into distinct channels of
finite size.  The decomposition of the full scattering problem
utilizes the generalized axial symmetry of the Hamiltonian $h$ in a Z
string background:
\be
[h, J_3 - n T_3 \gamma_5 ] = 0
\ee
where $J_3$ is the fermionic angular momentum,  $T_3$ is an isospin
generator and $n$ is the (integer) flux of the Z-string. Since
$\gamma_5$ does not commute with the free Hamiltonian $h_0$, we
have 
$[h_0, J_3 - n T_3 \gamma_5 ] \neq 0$. Thus, in contrast to standard
problems, the symmetries of the problem with background field are
\emph{not a subset} of the symmetries of the free problem, but
instead are incompatible with them. Thus, the setup for perturbation
theory (which considers the Hamiltonian $h_{\epsilon}=h_0+\epsilon V$)
has no axial symmetry \emph{at all}. The second problem is related
to the aforementioned IR problems of strings. Only special (gauge
invariant) combinations of Feynman diagrams of different perturbative orders
result in IR finite quantities. However, Born approximations
correspond to sums of Feynman diagrams of definite order in PT and
hence most likely will suffer from IR divergences. 

The case of D=2+1 is special. The only counterterm
that has a divergent co\-efficient --- $|\phi|^2$ --- is of a definite order
in PT. In this case we are able to use gauge invariance to argue that 
any background that has the same $|\phi|^2$ will suffer from the same
divergence and hence the Born approximation for some other \emph{fake}
background can be used to subtract the large-momentum tails
of the full phase shifts. So we can choose a background that has
$J_3$ as a symmetry generator but no IR divergences
\cite{Graham:2006}. 

In D=3+1 this procedure does not work because we now
have more counter terms with divergent coefficients that mix different
orders of PT. Furthermore, since there are additional constraints on
the profiles,  it is not clear how to
construct fake background solutions which have no winding and thus the
same symmetry as the free Hamiltonian -- if it's possible at all. 

At this point it is worthwhile to remember \emph{why} we want to
subtract Born approximations in the first place: First, we have to
subtract off the large-momentum behaviour of the summed phase shifts in order
to ensure UV convergence of the momentum integral. Second, with Born
approximations we know exactly what to add back in so that the overall
value of the determinant does not change, namely the corresponding
Feynman diagrams. 
Keeping these reasons in mind, we can extend our formalism
substantially: Instead of subtracting the
Born approximations of the original theory with the original background
fields, we can subtract off Born approximations from an arbitrary theory 
as long as a) we add back in the Feynman diagrams
from the same theory and b) that theory has the same divergences
(e.g., in dimensional regularization) as the original theory. 
In the remainder of this paper, we test this proposition.

\section{Fake subtractions \label{sec_fake}}
We consider two bosonic theories in D=3+1 that differ in the
string-like background potential they're coupled to: 
\bea
{\mathcal L}_{1,2}&=& \frac{1}{2} \partial_\mu
\phi\,\partial^\mu \phi - \frac{m^2}{2} \phi^2 - \frac{1}{2}
\sigma_{1,2} \,\phi^2  
\eea
with $\sigma_{1}(\vec{x}) = \alpha_1
\exp{\left(-\rho^2/w_1^2 \right)}, \sigma_{2}(\vec{x}) =
\alpha_2 \exp{\left(-\rho/w_2 \right)}$.
Demanding identical UV divergences requires
\bea
\int \rmd^3 x\, \sigma_{1}(\vec{x}) =  \int \rmd^3 x\, \sigma_{2}(\vec{x})\
\textrm{and} \ 
\int \rmd^3 x\, \sigma^2_{1}(\vec{x}) =  \int \rmd^3 x\,
\sigma^2_{2}(\vec{x}).
\eea
If our proposal of a fake Born subtraction works, the renormalized
vacuum polarization energy in (\ref{eqVacPolEn}) should be the same,
whether we use 
the original model 1 or its fake version 2 for the subtraction.
Since the bound states are not affected by the subtraction, it is
sufficient to consider the part
\bea
\hspace*{-5em} \tilde{E}_i &=&
   \frac{1}{2}\int \rmd k\, k \ln{\left( 1+ \frac{k^2}{m^2} \right)}
   \sum_{M} \frac{1}{\pi}\, \Big[\delta_{M} - \sum_{n=1}^2
   \delta_M^{(n)} (k; \sigma_i)  
       \Big] + \sum_{n=1}^2 E_{FD,ren}^{(n)}(k; \sigma_i) \label{eq_defTilde}
\eea
where the full phase shift is always computed in the original model 1,
while the subtraction is made with model 1 Born approximations
($\tilde{E}_1$) or with the fake model 2 Born approximations
 ($\tilde{E}_2$), respectively. Our proposal amounts to claiming that 
$\tilde{E}_1 = \tilde{E}_2$.  

It can be shown, using a Bessel function identity \cite{Leipzig}, that
$\sum_{M=0}^{\infty} \delta_M^{(1)}(k; \sigma_i) 
\propto \langle \sigma_i \rangle k^{q-2}$ where $q$ is the number of non-flat
spatial dimensions. By numerical calculation we can furthermore show
that
\bea
\hspace*{-7em} \frac{1}{2}\int \rmd k\,  k \ln{\left( 1+ \frac{k^2}{m^2}
  \right)}  \sum_{M} \frac{1}{\pi}\, \Big[\delta_M^{(2)}(k; \sigma_2)
- \delta_M^{(2)}(k; \sigma_1) \Big] 
      &=&  E_{FD,ren}^{(2)}(\sigma_2) -
      E_{FD,ren}^{(2)}(\sigma_1). \nonumber \\
      \eea
\begin{figure}
\centerline{
\includegraphics[width= \width, height = \height, angle = \picangle]{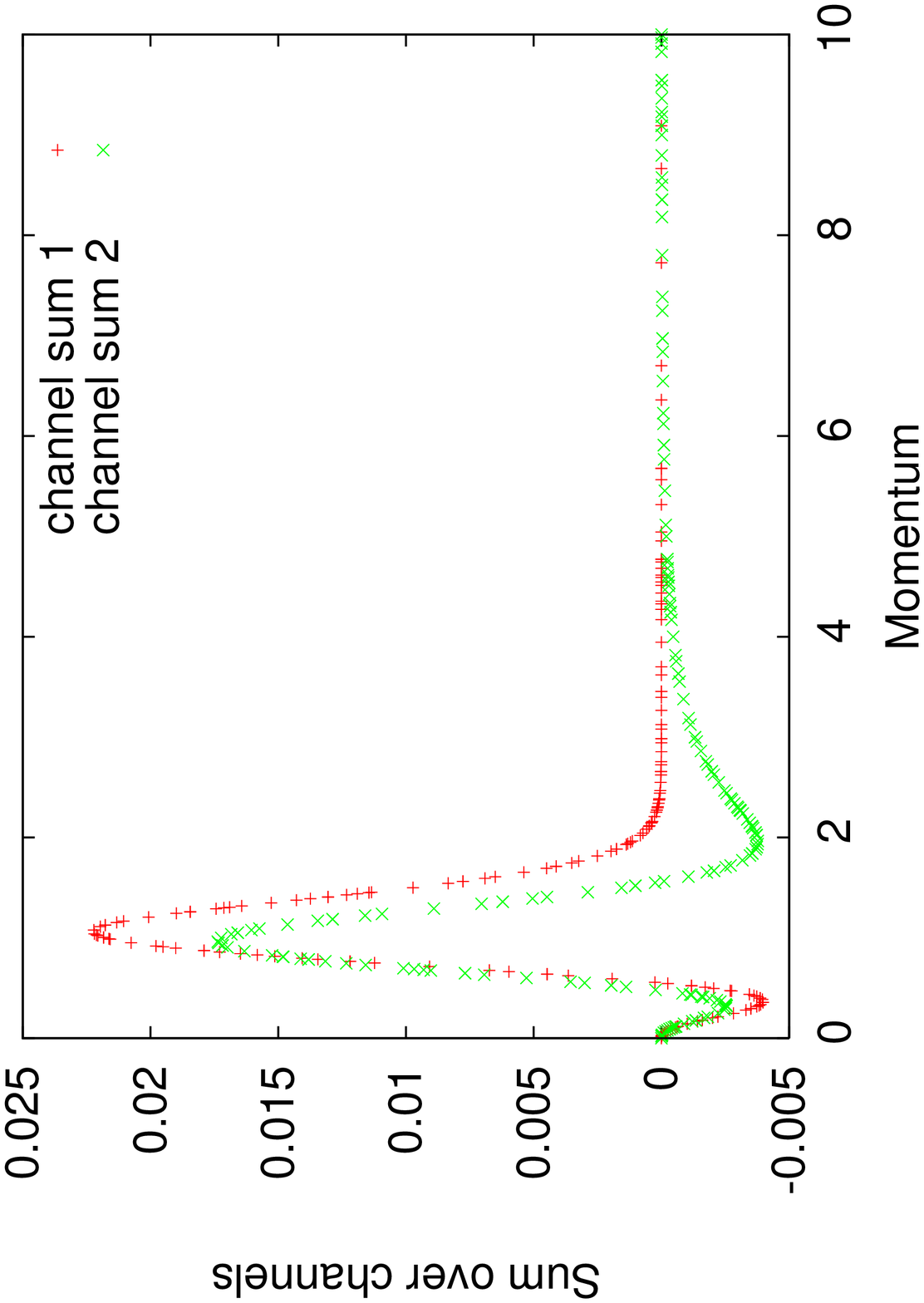} 
\hskip0cm
\includegraphics[width= \width, height = \height, angle = \picangle]{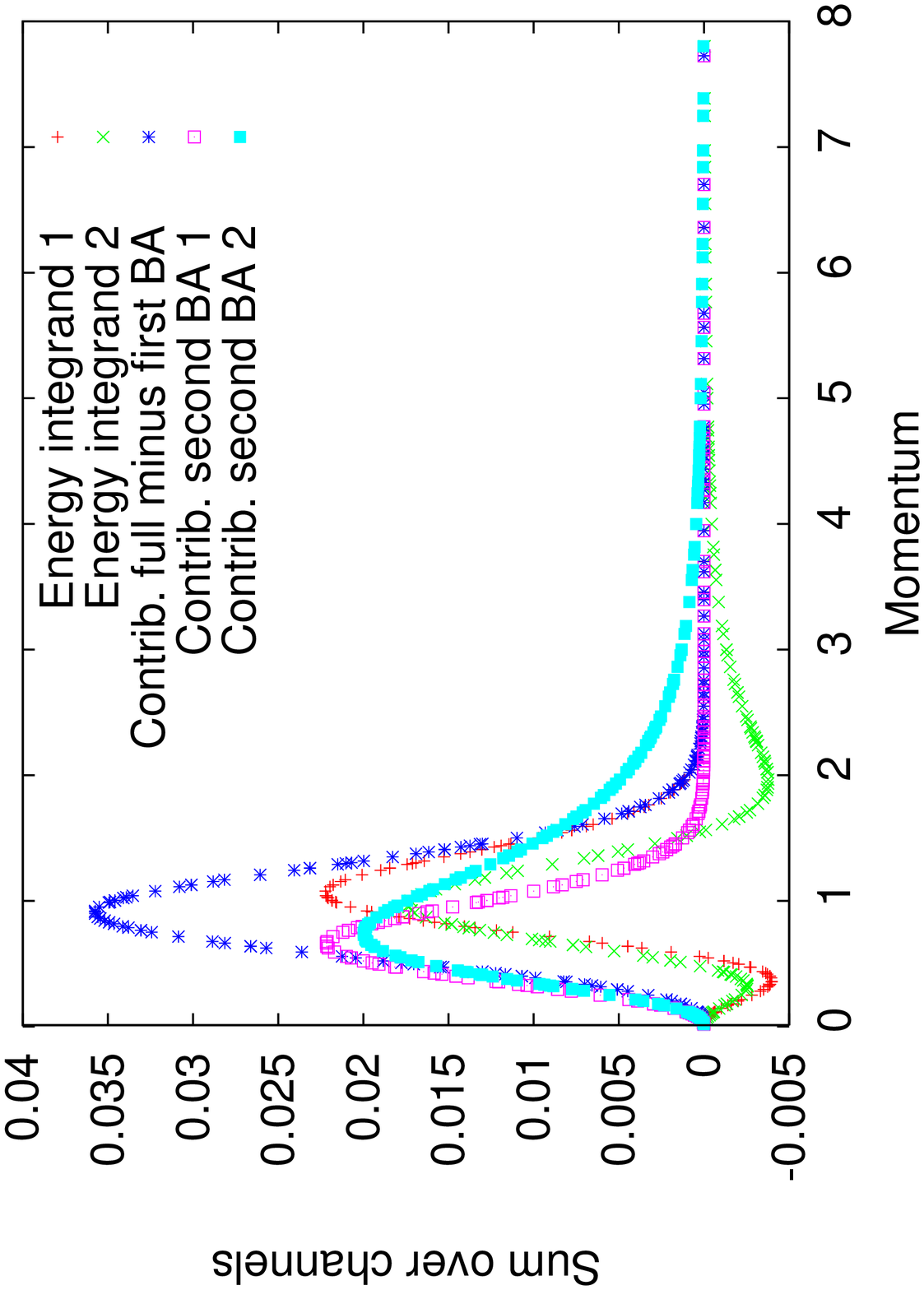}}
\leftline{~\hspace{-2.6cm}\begin{minipage}[l]{ \capwidth}
\caption{\label{figure1} \sf The 
left panel shows 
$k \ln{\left( 1+ k^2/m^2 \right)} 
     \sum_{M} \frac{1}{\pi}\, 
     \big[\delta_{M}(k; \sigma_1) - \delta_M^{(1)}(k; \sigma_1) -
     \delta_M^{(2)}(k; \sigma_i) \big]$ for $i=1,2$.
The right panel shows the decomposition furthermore into $k \ln{\left(
    1+ k^2/m^2 \right)} \sum_{M} \frac{1}{\pi}\, \big[ \delta_{M}(k;
\sigma_1) - \delta_M^{(1)}(k; \sigma_1) \big]$ and $k \ln{\left( 1+
    k^2/m^2 \right)} \sum_{M} \frac{1}{\pi}\, \delta_M^{(2)}(k; \sigma_i)$.}
\end{minipage}}
\end{figure}
This shows that as long as the renormalized determinant is finite,
both ways of subtracting Born approximations lead to the same result.
In the left panel of figure \ref{figure1} we see that the renormalized
determinant \emph{is}
going to be finite, since there is no large-momentum tail that might impede
the existence of the momentum integral.
The right panel, however, shows something troubling: 
Ordinarily, we would think that (\ref{eq_defTilde}) with just 
the first term of the Born series subtracted 
still contains the log-divergent 2$^{nd}$ order Feynman diagram. 
The Feynman series then suggests the same large $k$ behaviour for
the integrand of (\ref{eq_defTilde}) when the expression in
square brackets is replaced by the second order term of the Born
series. From the right panel of figure \ref{figure1}, one can clearly
see that this expectation is wrong: both contributions to the
integrand in (\ref{eq_defTilde}) fall off faster than $1/k^3$
individually.  
At first sight this indicates that the resulting
integral would be finite \cite{Pasipoularides:2000gg,
  Pasipoularides:2005gg}. This is in strong contradiction to the
fact the equivalent 2$^{nd}$nd order Feynman diagram is indeed
UV divergent.
As we will demonstrate in the rest of this paper the
catch comes from  an incorrect treatment of the non-uniformly
convergent sum and integral in (\ref{eq_defTilde}). 

\begin{figure}
\centerline{
\includegraphics[width= \width, height = \height, angle = \picangle]{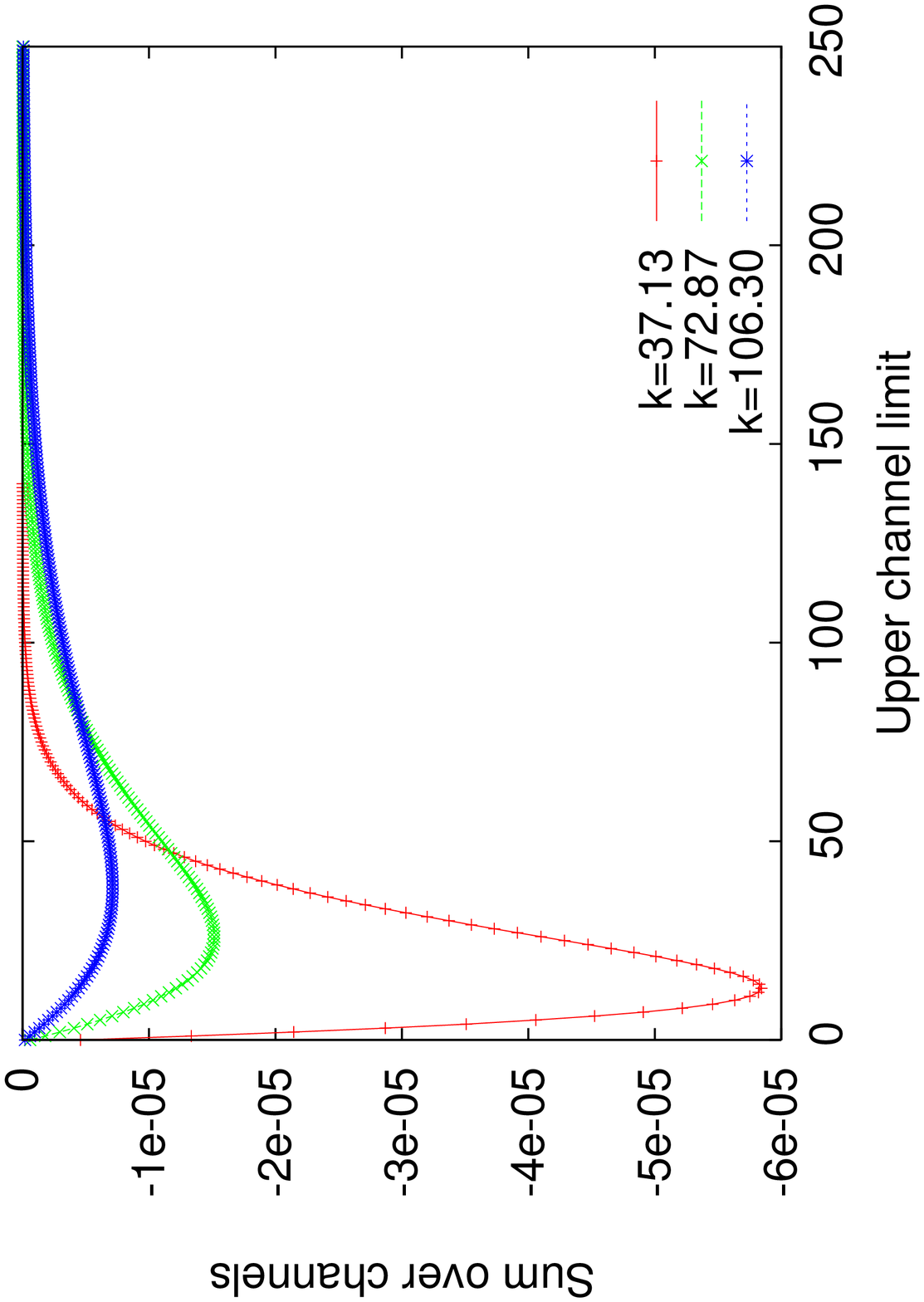} 
\hskip0cm
\includegraphics[width= \width, height = \height, angle = \picangle]{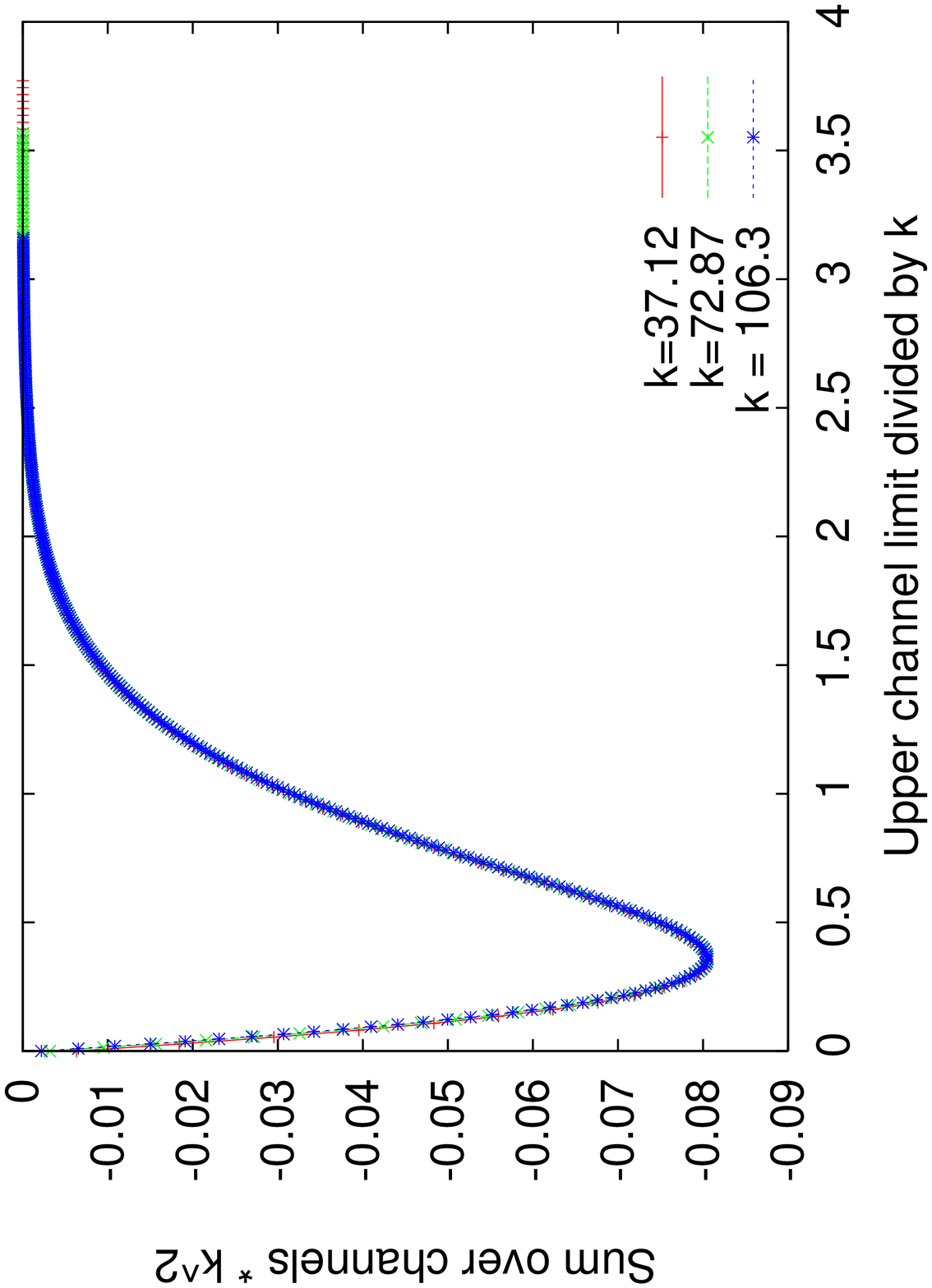}}
\leftline{~\hspace{-2.6cm}\begin{minipage}[l]{ \capwidth}
\caption{\label{figure2} \sf The 
left panel shows 
$S_N^{(2)}(k) = \sum_{M=0}^N \delta^{(2)}_M(k; \sigma_2)$ for
different values of $k$, the right panel shows $k^2 S_N^{(2)}(k)$
plotted vs. $N/k$. 
}
\end{minipage}}
\end{figure}

\begin{figure}
\centerline{
\includegraphics[width= \width, height = \height, angle = \picangle]{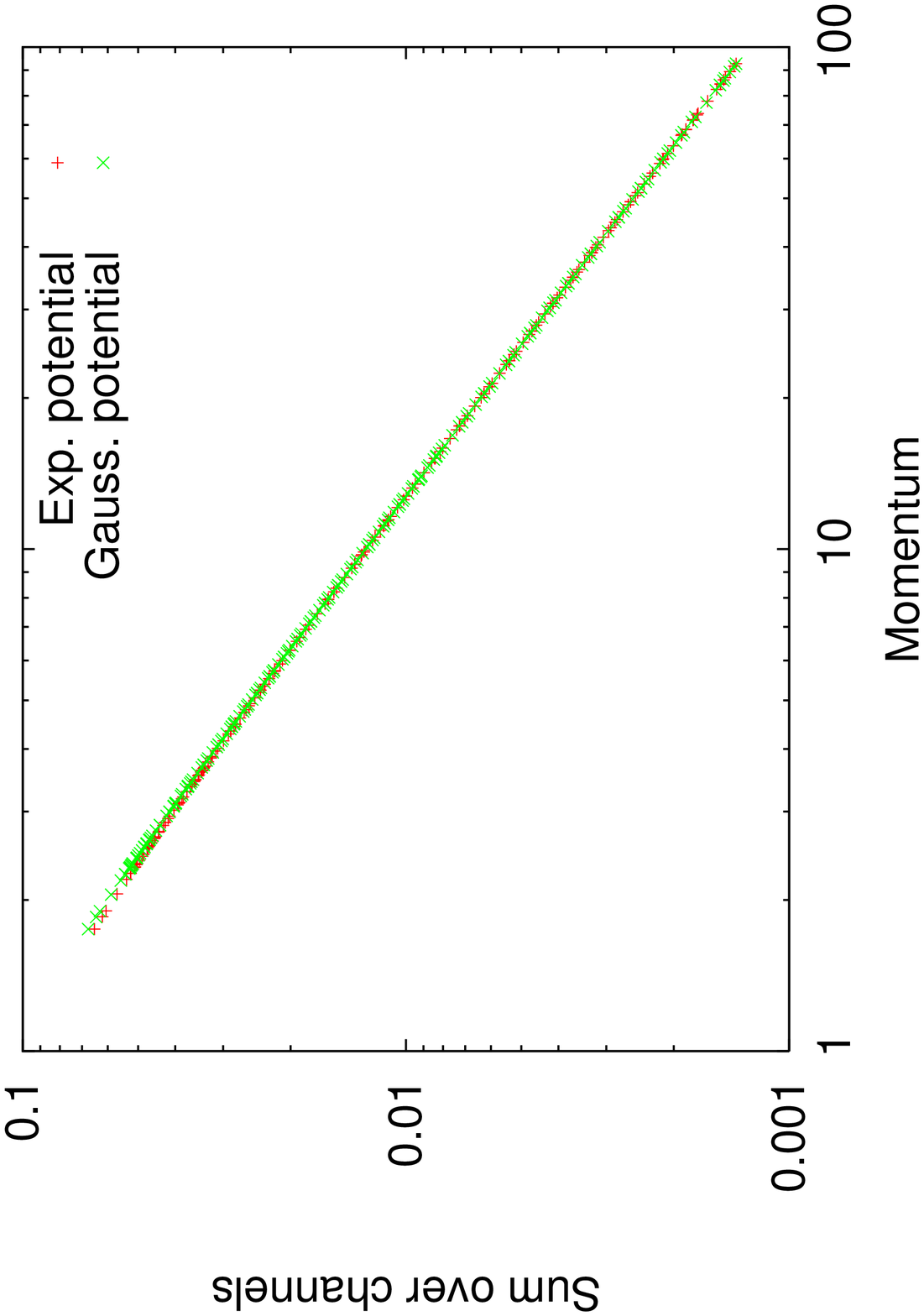} 
\hskip0cm
\includegraphics[width= \width, height = \height, angle = \picangle]{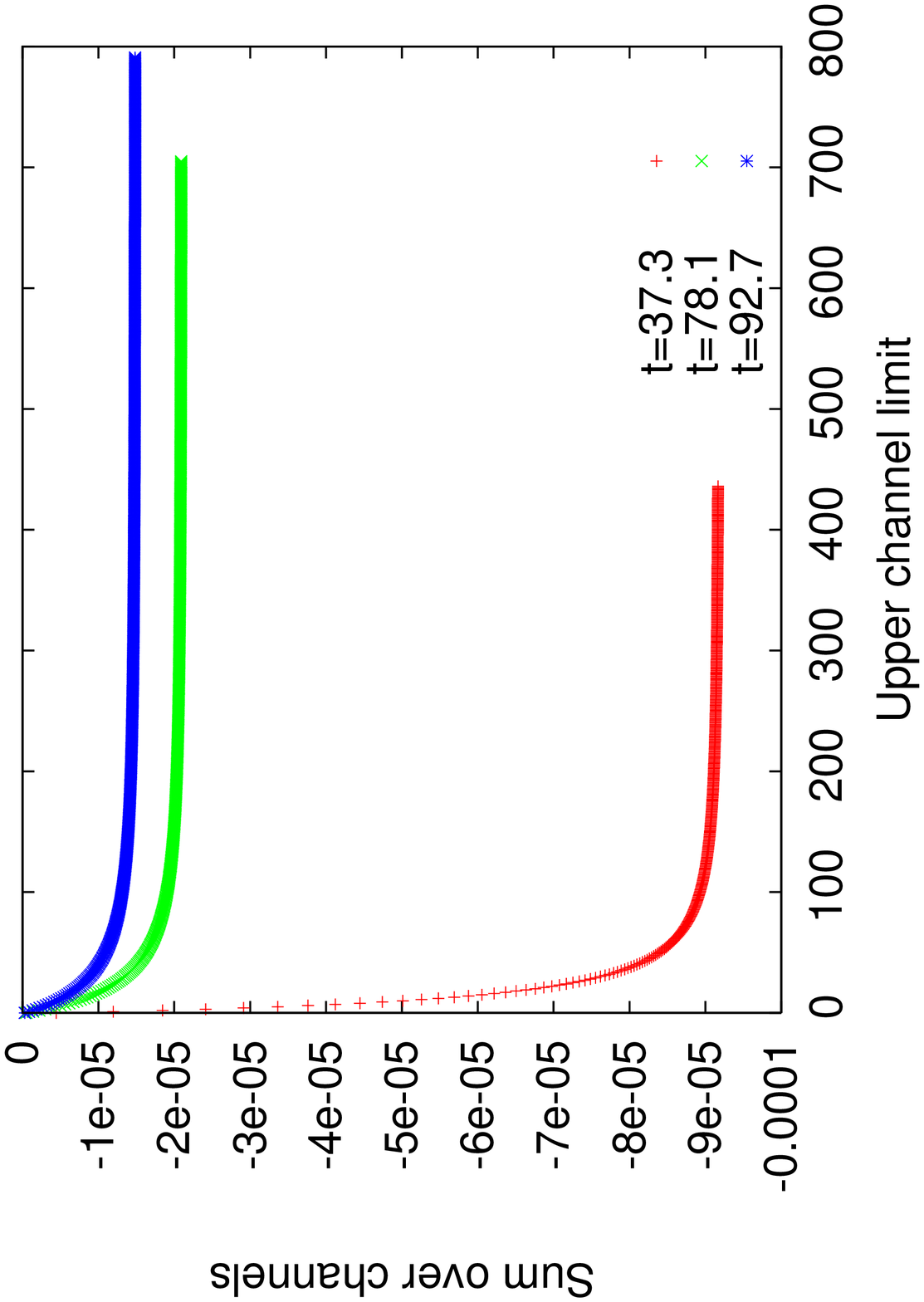}}
\leftline{~\hspace{-2.6cm}\begin{minipage}[l]{ \capwidth}
\caption{\label{figure3} \sf The left panel shows $t \sum_{M}
  \tilde{\delta}^{(2)}_M(t; \sigma_i)$ (which is the expression on the imaginary
  axis corresponding to $k \ln{\left( 1+ k^2/m^2 \right)} \sum_{M}
  \delta_M^{(2)}$ on the real $k$ axis) for different values of
  $t$ and $i=1, 2$. The right panel shows
  $\sum_{M=0}^N \tilde{\delta}^{(2)}_M(t; \sigma_2)$ plotted vs. $N$ for
  different values of $t$. 
}
\end{minipage}}
\end{figure}

First of all, it is worth noting that the second order Born
approximation phase shifts 
are not small by themselves, but they sum up to something
small, cf. figure \ref{figure2}.
Secondly, it is worthwhile to recall the formulation of functional
determinants in terms of scattering data \cite{Graham:2002fw,
  Graham:2002xq}. Strictly this is possible only in the upper half of
the complex k plane and not on the real k axis. Furthermore, that
derivation also suggests that instead of integrating over a channel
sum, one should integrate over momentum first to get a per-channel
contribution to the energy and then sum over channels. 
The analytic properties of scattering data ensure that the contribution
of a prescribed channel to the vacuum polarization energy can be identically
computed as an integral over real or imaginary momentum. We have
verified this result numerically. Of course, the corresponding
integrands are not expected to be identical. Consequently, the
momentum integrands obtained from summing over channels first are
expected to be different as well. 
The momentum integrand of the second order Born approximation phase
shift\footnote{More precisely one should talk about the imaginary
parts of the logarithm of the Jost function, because the concept of a
phase shift cannot be extended to the complex k plane.} 
on the imaginary axis shows exactly the
expected 1/t falloff\footnote{The momentum on the imaginary axis is
  denoted t, i.e., $k =i t$ with $t$ real.} for a logarithmically
divergent integral, cf. left panel of figure \ref{figure3}. In the
right panel, one can see that the 2$^{nd}$ order Born approximation on
the imaginary axis have the same sign for all channels and no
cancellations are present. Moreover, our general experience together
with our numerical investigations in this particular case suggest that the
sum over channels is uniformly convergent and we can freely
interchange the order of channel summation and momentum integral. 
\begin{figure}
\centerline{
\includegraphics[width= \width, height = \height, angle = \picangle]{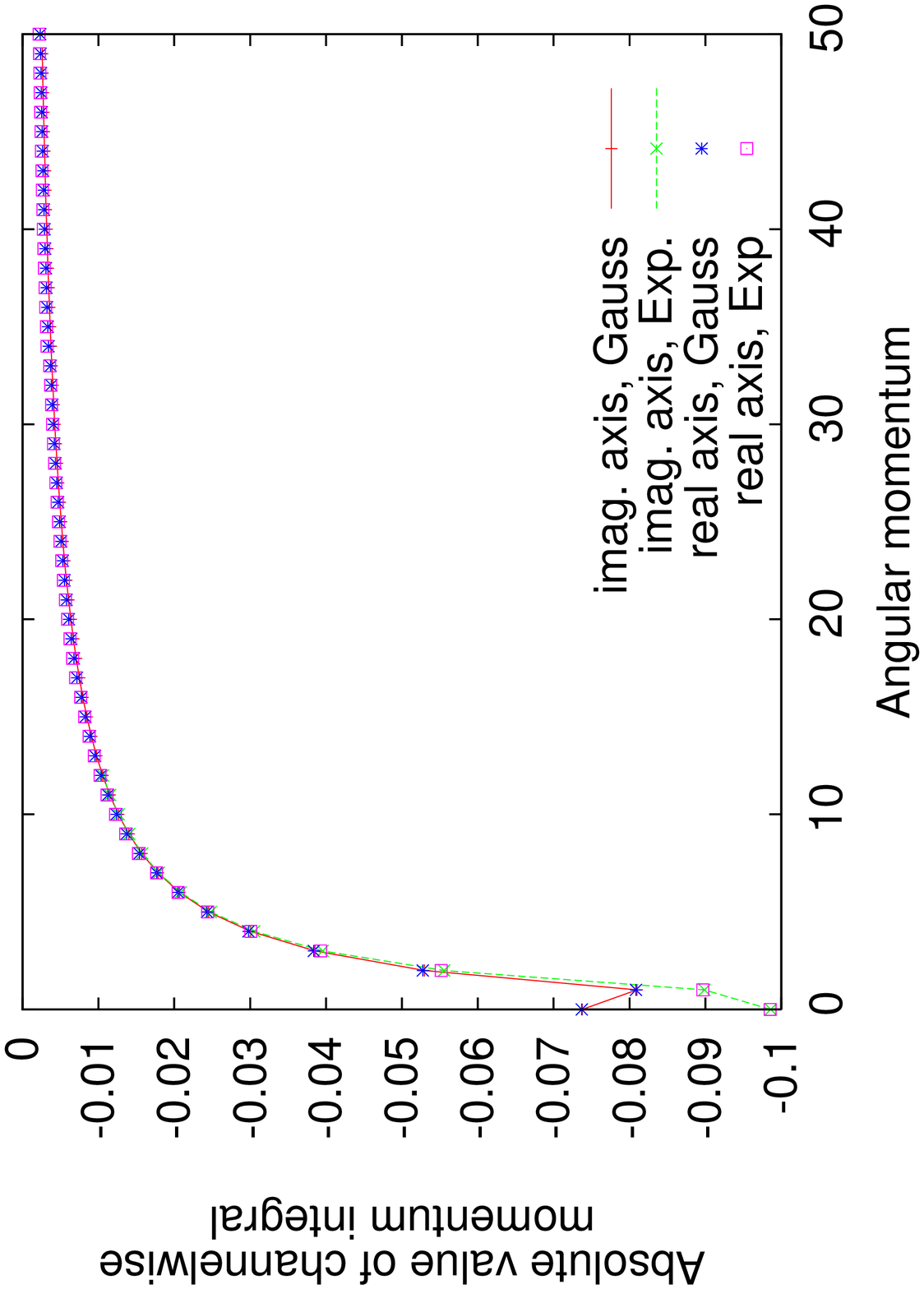} 
\hskip0cm 
\includegraphics[width= \width, height = \height, angle = \picangle]{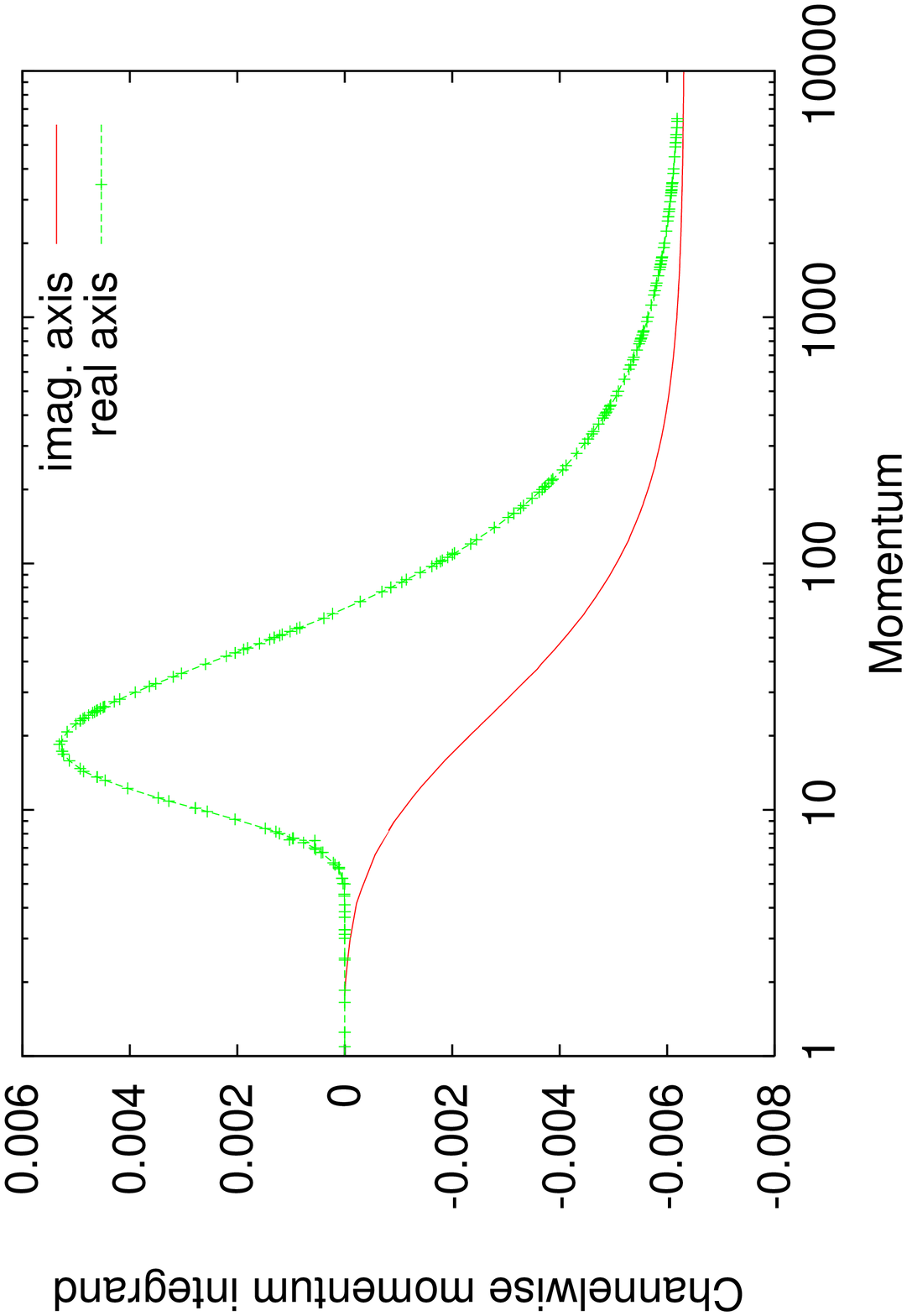}}
\leftline{~\hspace{-2.6cm}\begin{minipage}[l]{ \capwidth}
\caption{\label{figure4} \sf The left panel shows
$\int \rmd k\, k\, \ln{(1+k^2/m^2)} \delta_M^{(2)}(k)$ - denoted as 'real
axis' -  and  $\int \rmd t\, t\, \tilde{\delta}_M^{(2)}(t)$ - denoted as
'imag. axis' - for both $\sigma_1$ and $\sigma_2$ as potentials. Note
that for larger values of $M$, the curves for $\sigma_{1,2}$ coincide
and show the expected $1/M$ fall off. The right panel shows the
momentum integrand for both real and imaginary axis for channel $M=10$.}
\end{minipage}}
\end{figure}

The left panel of figure \ref{figure4} shows clearly that the problem on
the real axis is the interchange of integration and summation, because
by integrating first over momentum and then summing over channels we obtain the
expected logarithmic divergence, whereas in the other order 
the result is apparently finite. The right panel shows how the
(identical) values on the real and imaginary axis are obtained --- on
the imaginary axis we see a monotonic approach to the final value, while on
the real axis we see a zero crossing. 
This zero is not accidental, but rather a consequence of the
fundamental sum rule 
\bea
\int \rmd k\, k\, \delta^{(2)}_l(k) = 0, \label{eq_sumrule}
\eea
proved in \cite{Graham:2001iv}. 
On the imaginary axis no corresponding sum rule exists. 
\begin{figure}
\centerline{
\includegraphics[width= \width, height = \height, angle = \picangle]{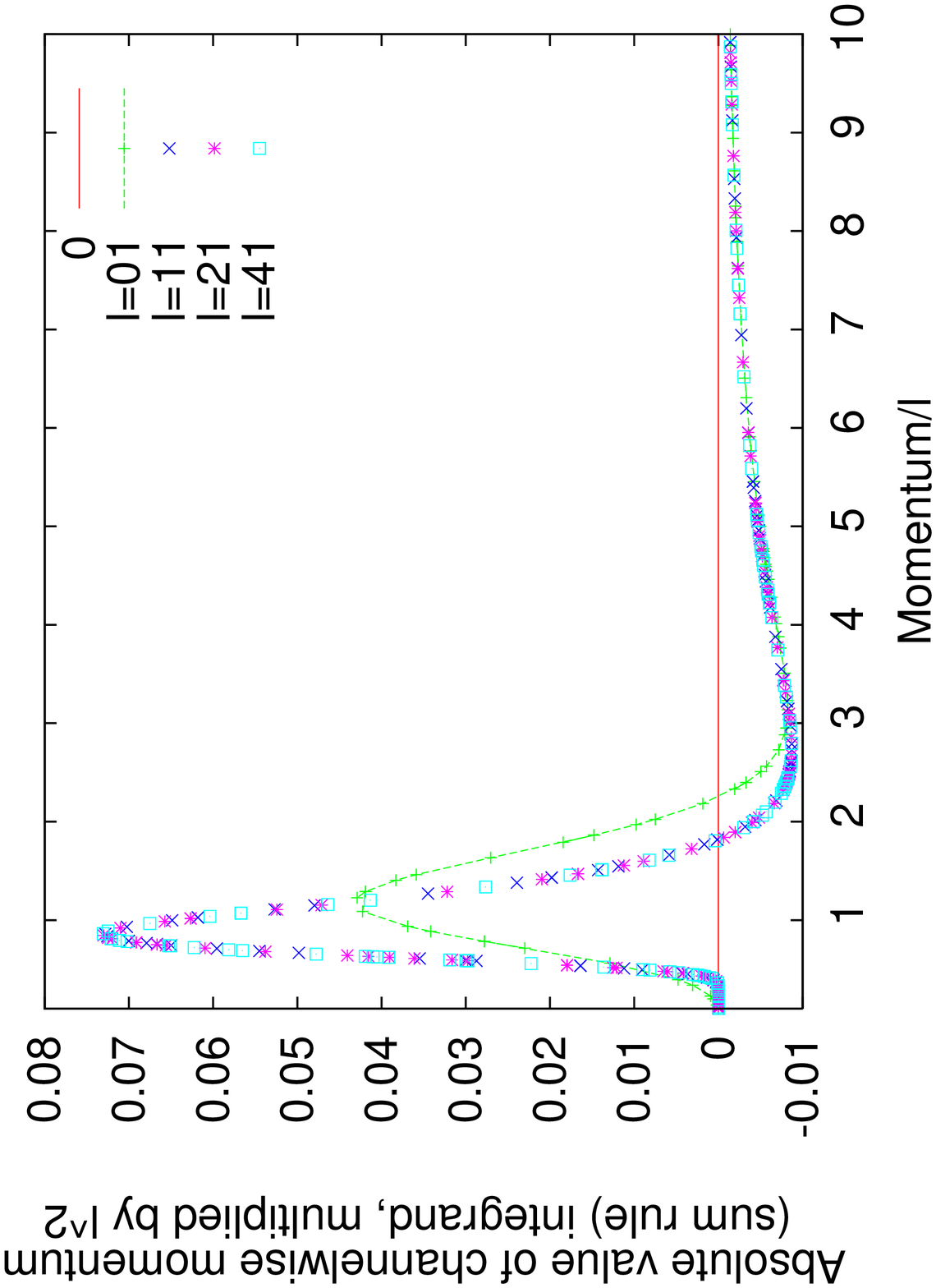} 
\hskip0cm
\includegraphics[width= \width, height = \height, angle = \picangle]{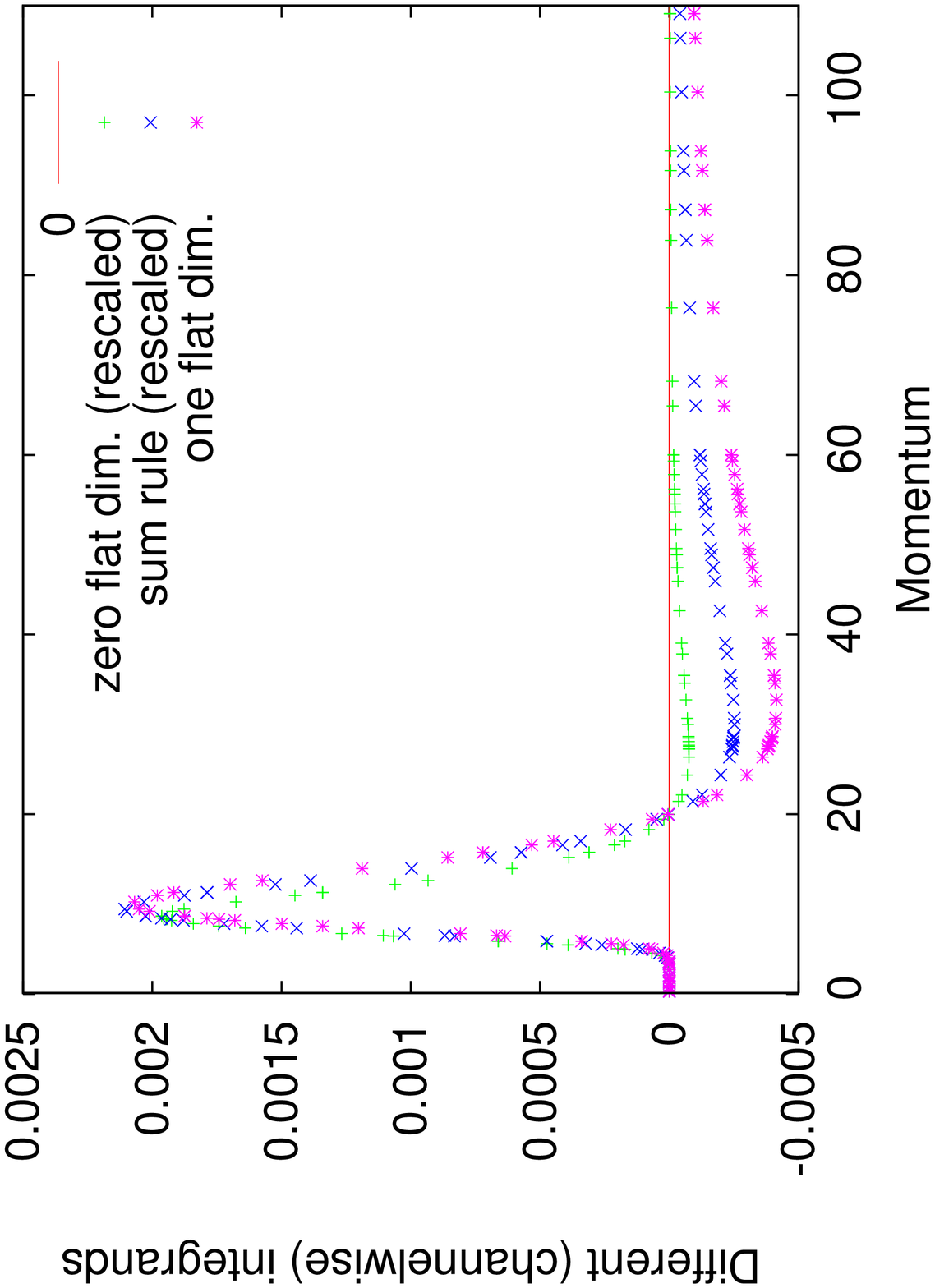}}
\leftline{~\hspace{-2.6cm}\begin{minipage}[l]{ \capwidth}
\caption{\label{figure5} \sf The left panel shows another scaling
  rule; the most important point is that the zero crossing moves
  outwards on the real $k$ axis as $M$ increases. The right panel 
  compares the integrands of the per-channel contribution to the energy
  with zero and one flat dimension to the integrand of the sum rule.
}
\end{minipage}}
\end{figure}
The consequence of this zero crossing can be seen in figure \ref{figure5}.
From the right panel of this figure one can see clearly that
while
for zero flat dimensions the dominant contribution to the per-channel
contribution of the energy comes from the low-k region, for one flat
dimension the dominant contribution comes from the region $k \cdot w >
M$, where $w$ is the characteristic width of the potential. This
result explains why the interchange of summation
and integration fails on the real axis for one flat dimension (and not
for zero);  if one integrates first over momentum, one
obviously is able to catch the dominant large momentum
contribution. 
If one sums over channels first, however, one finds that 
for $M>k \cdot w$ the phase shifts drop off exponentially. Hence it seems
sufficient to sum up to $M_{\rm max}\approx(2 \ldots 3) k \cdot
w$. Since the momentum integral will be terminated at some finite
value $k_{\rm max}$, this procedure obviously misses important contributions
from channels with $M>M_{\rm max}$. 
Hence, integration and summation can not be
interchanged\footnote{During our investigation of the second 
order Born approximation phase shifts, we noted that $k^3
\delta^{(2)}_M(k)$ is not a 
function of $k$ and $M$ individually, but is just a function of $k/M$ (plus
corrections, but those are very small for $M>1$). This
allows to map $\sum_{M} \delta^{(2)}_M(k)$ via the (leading order)
Euler-Maclarin 
formula to $\int \rmd k\, k\, \delta^{(2)}_M(k)$ which might ultimately explain
why on the real axis the sum over channels is so small. For definite
conclusions it will be necessary to investigate the next-to-leading
orders, though.}. Another piece of evidence is that the momentum
integrand obtained from summing over channels first is --- as far as we
can tell from our numerical investigations --- strictly
positive for any prescribed finite momentum, which is in direct
contradiction to the sum rule from 
(\ref{eq_sumrule}).

\section{Conclusions\label{sec_conclusions}}
We have shown that for finite quantities one can successfully replace
Born approximations in one theory by Born approximations from another
as long as the Feynman diagrams show the same divergences. This is a
major improvement over existing formalisms:  One does not have to deal
with problems that originate from non-vanishing structures at spatial
infinity, but still one knows exactly what to add in
again to implement renormalization conditions known from
perturbation theory. In these
conference proceedings, we have focused on a pathological case where
the ultimate object of interest does not strictly speaking exist (as
it is UV divergent); but the apparent convergence of the
vacuum polarization diagram in the phase shift formulation was important to
understand. This (erroneous) convergence only
occurs when formulating the vacuum polarization energy in terms
of scattering data for real momenta. Analytically continuing to
the imaginary axis yields the UV structure that is consistent
with the analysis of the Feynman diagrams without any further subtleties
like ordering of limits. Though this favors the
formulation in terms of imaginary momenta it contains drawbacks
that we did not go into --- one needs to sum a lot more channels and it
\emph{may} pose challenges for fermions. 

\ack
One of us (OS) would like to thank
the organizers of QFEXT'07 for the marvelous job they have done in
making the conference happen and the generous financial support they provided
and the other participants for many interesting conversations. 
NG was supported by the National Science Foundation (NSF)
through grant PHY-0555338, by a Cottrell College Science Award from
Research Corporation, and by Middlebury College.

\end{document}